\documentclass[12pt]{JHEP3}
\usepackage{amsmath,amssymb,graphicx}

\title{Entanglement and Nonunitary Evolution}
\author{Ram Brustein \\ Department of Physics, Ben-Gurion University,
    Beer-Sheva 84105, Israel \\ E-mail: ramyb@bgumail.bgu.ac.il}
\author{Martin B. Einhorn \\ Kavli Institute for Theoretical
    Physics, University of California, Santa Barbara, CA 93106-4030.
    \\ Michigan Center for Theoretical Physics, Randall Laboratory, The
    University of Michigan, Ann Arbor, MI 48109-1120 \\
    E-mail: meinhorn@kitp.ucsb.edu}
\author{Amos Yarom \\ Department of Physics, Ben-Gurion University,
    Beer-Sheva 84105, Israel. \\
    Arnold--Sommerfeld--Center for Theoretical Physics, Department f\"ur Physik,
    Ludwig--Maximilians--Universit\"at M\"unchen, Theresienstra\ss e 37, 80333
    M\"unchen, Germany.
    \\ E-mail: yarom@theorie.physik.uni-muenchen.de }
\abstract{We consider a collapsing relativistic spherical shell for
a free quantum field. Once the center of the wavefunction of the
shell passes a certain radius $r_s$, the degrees of freedom inside
$r_s$ are traced over. We show that an observer outside this region
will determine that the evolution of the system is nonunitary. We
argue that this phenomenon is generic to entangled systems, and
discuss a possible relation to black hole physics.}

\preprint{MCTP-06-22, NSF-KITP-06-67, LMU-ASC 59/06}
\begin{document}

%%% ----------------------------------------------I------------------------
\section{Introduction}
Consider a free massless scalar field in a pure state $|\psi\rangle$
describing a spherically symmetric inward collapsing shell. Imagine
that once the shell reaches a certain radius $r_s$, an observer
located at $r>r_s$ has no access to the region $r<r_s$. This implies
that the state seen by this outside observer is a mixed state
described by the density matrix  $\rho_{out} =
\text{Tr}_{r<r_s}|\psi\rangle\langle \psi|$. We shall investigate
the time evolution of the state $\rho_{out}$.

There are general arguments, which we discuss below, that show that
a generic entangled state will seem to evolve in a nonunitary way if
part of the Hilbert space is inaccessible---a property that leads to
dissipation effects in many body systems. In this paper, we shall
show that the eigenvalues of $\rho_{out}$ are time dependent.  This
implies that $\rho_{out}$ does not evolve in a unitary manner (see
figure \ref{F:mainplot}).  To see this, consider a quantum system
prepared in an initial state defined by the density matrix
$\rho(0)$. If the evolution is unitary, then the density matrix
$\rho(t)$ describing the system at a later time is related to
$\rho(0)$ by $\rho(t) = U^{\dagger}(t)\rho(0)U(t) $. Since this is a
similarity transformation, the eigenvalues of $\rho(t)$ must be the
same as those of $\rho(0)$, and are consequently  time-independent.
Conversely, consider a density matrix $\rho(t)$ whose eigenvalues
are time dependent.  In this case, for the same reason, a unitary
operator $U(t)$ such that $\rho(t) = U^{\dagger}(t)\rho(0)U(t)$ does
not exist.  Therefore, the time evolution of a system will be
unitary if and only if the eigenvalues of $\rho$ are time
independent.

Previous investigations of the relationship between entanglement,
entropy and area \cite{Srednicki} have shown that the von~Neumann
entropy $S=\text{Tr}_{r>r_s} (\rho_{out} \ln\rho_{out})$ of an
initial vacuum state $|0\rangle$ is proportional to the surface area
of the region $r<r_s$, suggesting a connection with black hole
entropy. This observation (see also \cite{Bombelli}) has been
studied in the literature from various perspectives
\cite{Sol,Casini1,Fursaev1,Das,Myung,Buniy1,Parikh,Dou,Plenio,Areascaling,Casini2,Einhorn,Implications,coldbosons,Holred,Mukohyama,Holzhey}
(see \cite{AYThesis,Sorkin} for a review). In flat space, it can be
shown that the entropy (and other thermodynamic quantities) will
scale as the surface area of the inaccessible volume even when it is
nonspherical \cite{Areascaling,Implications,Casini2}. This area
dependence may be related to thermodynamic properties of Unruh
radiation \cite{TandA,KabStr}.

The flat space arguments relating entanglement entropy and area may be extended to a black hole background \cite{Israel,Iorio,EntinST}.
Since this work is key in understanding the relation between our results to black hole evolution, we shall briefly review it. Consider, for simplicity, an eternal uncharged and non rotating black hole. This black hole has two asymptotic regions separated by a horizon. As a result of this causal structure, an observer in one asymptotic region has no access to the degrees of freedom in the other asymptotic region. In contrast, the global vacuum of this system (the Hartle-Hawking vacuum) extends to both asymptotic regions. An observer in one asymptotic region will observe the state obtained from the vacuum after degrees of freedom in the other asymptotic region are traced over---the degrees of freedom ``behind'' the horizon. The resultant state is not a pure state but rather a mixed state which turns out to be precisely the density matrix seen by an observer restricted to one asymptotic region at the Hawking temperature. As in the flat space case, the entropy associated with this state is proportional to the surface area of the horizon. This mechanism, which is a result of the entangled nature of the vacuum state, allows one to interpret the black hole entropy as coming from entanglement, an interpretation which is consistent with the string theory evaluation of black hole entropy \cite{EntinST,MalStr,HawMalStr,LouMar,MarYar} (see also
\cite{entAdSCFT1,entAdSCFT2,entAdSCFT3,entAdSCFT4,entAdSCFT5,entAdSCFT6,entAdSCFT7}
for a recent discussion of black hole entropy and entanglement in
the context of AdS/CFT.)

Thus, following \cite{Srednicki}, the current work is a natural next
step in relating features of entanglement to black hole physics.
While it is generally believed that black holes have entropy
proportional to their surface area \cite{Bekenstein} there is, at
the moment, some debate regarding their time evolution.

In \cite{Hawkingnonu} it was argued that the evolution in time of a
black hole is nonunitary (see
\cite{Mav1,Mav2,Mav3,Giddings1,Giddings2,Vijay,Horowitz2,Einhorntalk,Horowitz1,Hawkingrecent,Hsu}
for recent discussions and implications). This is a troubling
feature of black holes since it clashes with our understanding of
quantum mechanics. For instance, if the evolution is nonunitary a
pure state (for which the density matrix has a single nontrivial
eigenvalue) may evolve into a mixed state. From the entanglement
point of view this behavior is not in contradiction with the
principles of quantum mechanics or any laws of physics, rather it is
a simple consequence of the entangled nature of the system. If one
does not have access to part of the Hilbert space then probability
flowing into the inaccessible region will seem to disappear leading
to nonunitary evolution.
%%%
%%%
\section{General arguments}
%%%
%%%
There are several approaches one may take when trying to exhibit the
nonunitary evolution of an entangled system. Many of these have been
studied in the context of dissipation of energy from a system to a
heat bath (see for example \cite{dissipation} and references
therein) and may be directly applied to our discussion. We review
them here for completeness.

We would like to exhibit the nonunitary evolution of an initial pure
state $\rho(0)=|\psi\rangle\langle\psi|$ seen by an observer who has
no access to a certain region of space. We evolve our initial state
in time using the Liouvillian operator
$\mathcal{L}=-\imath[H,\quad]$, so that $\dot{\rho}(t) = \mathcal{L}
\rho(t)$.

We shall split our Hilbert space of states into two parts, which we
will refer to as the `in' region and the `out' region, such that the
Hamiltonian can be written as $H=H_{out} + H_I + H_{in}$ with $H_I$
an interaction term.   Since $\mathcal{L}$ is linear, this
correspondingly results in
$\mathcal{L}=\mathcal{L}_{out}+\mathcal{L}_{I}+\mathcal{L}_{in}$.

Consider an observer restricted to, say, the `out' region of space,
and let $R$ be an operator that restricts the density matrix to this
region.  This corresponds to
 ``integrating out" the unobservable degrees of freedom.  (In the path integral formalism,
 $R$ is given by eq.~(\ref{E:rhoinfield}) below.)
We assume that the `in' and `out' regions are fixed, so that $R$ is
time-independent (though a generalization to a time dependent $R$ is
straightforward).

Suppose that at some time $t=0$ the state of the observer is given
by $\rho_{out}(0)=R \rho(0)$. The state $\rho_{out}$ will then
evolve in time according to $\dot{\rho}_{out}(t) = R \dot{\rho}(t)$.
Note that, once we restrict observables to the `out' region, a
further restriction will not change the state of the system,
therefore $R^2=R$.

To find $\dot{\rho}_{out}$, we act, in turn, with $(I-R)$ and $R$ on
$
%\label{E:dotrhoT}
    \dot{\rho} = \mathcal{L} \rho_{out} + \mathcal{L}(I-R)\rho
$, %
obtaining equations for $\dot{\rho}_{out}$ and $(I-R) \dot{\rho}$.
Plugging the (formal) solution to $(I-R)\rho$ into the equation
for $\rho_{out}$ we obtain the Nakajima-Zwanzig
\cite{Nakajima,Zwanzig,boulder} form of the master equation
\begin{align}
\notag
    \dot{\rho}_{out}(t) =& \left( R\mathcal{L}\right)
    \rho_{out}(t)
    +
    \int_0^t \left( R \mathcal{L}  \right)\big|_t
    \exp\left[\int_{t-\bar{t}}^t
    \left((I-R)\mathcal{L}\right)
    dt^{\prime}\right]
    \left((I-R)\mathcal{L}\right) \rho_{out}\big|_{t-\bar{t}} d\bar{t}+ \\
\label{E:rhoout}
    &+
    \left( R \mathcal{L} \right)\big|_t \exp\left(\int_0^t (I-R)\mathcal{L}
    dt^{\prime\prime}\right)(I-R)\rho(0).
\end{align}

To simplify (\ref{E:rhoout}) we note that the restriction $R$
commutes with time evolution in the `in' and `out' regions
$[R,\mathcal{L}_{out}]=[R,\mathcal{L}_{in}]=0$. Also, we assume
that, $\mathcal{L}_{in} R = 0$. Finally, since $
    R \exp\left[\int_{t-\bar{t}}^t
    \left((I-R)\mathcal{L} \right)
    dt^{\prime}\right]
    \left((I-R)\mathcal{L}_I\right)(t-\bar{t})
    =0
$, %
we find
\begin{align}
\notag
 \dot{\rho}_{out}(t) =& R(\mathcal{L}_{out} +\mathcal{L}_I)
    \rho_{out}(t)+
    \\
\notag
    &+
    \int_0^t \left( R \mathcal{L}_I\right)\big|_t
    \exp\left[\int_{t-\bar{t}}^t
    \left((I-R)\mathcal{L} \right)
    dt^{\prime}\right]
    \left((I-R)\mathcal{L}_I\right) \rho_{out}\big|_{t-\bar{t}} d\bar{t} +\\
\label{E:rhoin2}
    &+
    R \mathcal{L}_I(t) \exp\left(\int_0^t (I-R)\mathcal{L}
    dt^{\prime\prime}\right)(I-R)\rho(0).
\end{align}

The first term in eq.~(\ref{E:rhoin2}) is the standard term
that one expects for unitary time evolution of the state seen by the
observer in the `out' region. The second term represents the (time
retarded) oscillations flowing into the `out' region from
 the `in' region that is being traced over (``beyond the horizon'').
 The last term represents contributions leaking out of the inaccessible
 region from those initially present.  In most many-body applications, it is arranged
 or assumed that $(I-R)\rho(0)\approx 0,$ so that this term can be neglected.

The interaction term in the Hamiltonian, $H_I$, is essential for
generating the nonunitary evolution. In a field theory setting, this
interaction term includes the spatial derivative coupling fields
across the boundary.  Thus, it exists even for a free field theory.

%%% ------------------------------------------------------------------------
\section{Explicit construction of nonunitary evolution.}

We wish to explicitly exhibit the nonunitary evolution measured by
an observer who does not have access to part of a system.
Consider the time evolution of an inward collapsing spherical
shell of a free massless field in $3+1$ dimensions from the point
of view of an observer who does not have access to the region
$r<r_s$. We choose such a configuration for two reasons: (a) to
make suggestive contact with gravitational collapse, and (b) since
any choice of an initial state that is an eigenstate of the
Hamiltonian (such as the vacuum state) will not exhibit any time
evolution due to its stationary nature.

Let $\Psi_k[\phi(\vec{x})]$ be the wavefunctional of a single
particle with energy $E_k = |\vec{k}|$ of a free massless scalar
field $\phi(\vec{x})$ in three space dimensions. An arbitrary
configuration $\Psi$ for such a particle is specified by $\Psi =
\sum_k f_k \Psi_k$. We construct a collapsing shell by choosing $\{
f_k \}$ such that only s-wave modes are excited and such that the
initial position space wavefunction is centered at some radius $r_i$
and is collapsing toward the origin. To obtain $f_k$ explicitly,
consider the wavefunction
\begin{equation}
\label{E:wfcontinuum}
    |\psi\rangle = A^{-1/2}\int \hat{f}(r) \phi(\vec{x})|0\rangle
        d^3x
\end{equation}
where $A$ is a normalization constant determined by the condition
$\langle \psi | \psi \rangle = 1$ and $\hat{f}(r)$ is real. The
average momentum of the shell, $p$, is set by substituting
$\hat{f}(r) \to e^{\imath p r}\hat{f}(r)$. The set $\{f_k\}$ may be
obtained from the expansion of the field $\phi$ in terms of ladder
operators.

The density matrix of the region seen by an observer in the exterior
is given by
\begin{equation}
\label{E:rhoinfield}
    \rho_{out}(\phi_{out},\phi^{\prime}_{out})
    =
    \int
        D\phi_{in}
        \Psi(\phi_{in},\phi_{out}) \Psi(\phi_{in},\phi^{\prime}_{out})^{\dagger}.
\end{equation}
A similar density matrix was constructed in a somewhat different
context in \cite{Cardy}.

As in \cite{Srednicki,Das}, one can calculate the functional
integral (\ref{E:rhoinfield}) by discretizing space on a radial
lattice. We expand the scalar field $\phi$ in partial wave
components
\begin{equation}
    \phi_{l,m}(r) = r\int Z_{lm}(\theta,\varphi)
    \phi(\vec{x})d\Omega, \qquad
    \pi_{l,m}(r) = r\int Z_{lm}(\theta,\varphi)
    \pi(\vec{x})d\Omega,
\end{equation}
where $r=|\vec{x}|$ and $Z_{0,0}=Y_{0,0}$, $Z_{l,m}=\sqrt{2}
\text{Re} Y_{l,m}$ for $m>0$, and $Z_{l,m}=\sqrt{2} \text{Im}
Y_{l,m}$ for $m<0$. $Z_{l,m}$ are orthonormal and complete. This
allows us to write the Hamiltonian as $H=\sum H_{l,m}$ with
\begin{equation}
\label{E:Hamcont}
    H_{l,m} = \frac{1}{2} \int_{0}^{\infty}dr
        \left[
            \pi_{l,m}^2(r)+r^2 \left(
                \partial_r \left(
                    \frac{\phi_{l,m}(r)}{r}
                        \right)
                \right)^2
            +\frac{l(l+1)}{r^2} \phi_{l,m}^2(r)
        \right].
\end{equation}

We discretize the radial coordinate on a lattice of spacing $a$ and
of size $L=(N+1)a$ with boundary conditions such that the fields
vanish at $r=L$. The expressions for the $H_{l,m}$'s
(\ref{E:Hamcont}) become those of Hamiltonians of coupled harmonic
oscillators.
\begin{align}
\notag
    H_{l,m}&=\frac{1}{2a}\sum_{j=1}^{N}
        \left[\pi_{l,m,j}^2+
            (j+\frac{1}{2})^2
            \left(\frac{\phi_{l,m,j}}{j}-\frac{\phi_{l,m,j+1}}{j+1}
                \right)^2
            +\frac{l(l+1)}{j^2} \phi_{l,m,j}^2
        \right]\\
\label{E:Hlm}
    &\equiv \frac{1}{2a}\left(
        \pi^2 + \phi K_{l,m} \phi \right).
\end{align}
Note that all the oscillators are coupled due to the spatial
derivatives of the field. This coupling generates the interaction
Hamiltonian described earlier.

The wavefunction for a single particle shell is constructed as in
the continuum (\ref{E:wfcontinuum})
\begin{equation}
    |\psi\rangle =
        \frac{a^2}{A^{1/2}}  \sum_{i=1}^N i \hat{f}_i \phi_{0,0,i} |0\rangle.
\end{equation}
We wish to write this in a single particle eigenbasis of the
Hamiltonian $\{ a_i^{\dagger}|0\rangle \}_{i=1}^N$. Let $y_i =
S^0_{ij} \phi_{0,0,j}$, where $S^0$ diagonalizes $K_{0,0}$ defined
in (\ref{E:Hlm}). We find that
\begin{align}
\label{E:psiini}
    |\psi\rangle &=
        A^{-1/2} \frac{1}{\sqrt 2} \sum_{j,k=1}^N j \hat{f}_j S^0_{kj}
        (\Omega^{0}_{k,k})^{-\frac{1}{2}} a_k^{\dagger}
        |0\rangle
\intertext{with}
\notag
    A&=\frac{1}{2}\sum_{j,j^{\prime}} j^{\prime} \hat{f}_{j^{\prime}} \left(K_{0,0}^{-1/2}\right)_{j^{\prime} j} j
    \hat{f}_j.
\end{align}
In (\ref{E:psiini}) we have defined $(\Omega^0)^2 = S^0 K_{0,0}
S^{0\,T}$. As in the continuum, one can generate a collapsing shell
by substituting $\hat{f}_j \to e^{\imath k j} \hat{f}_j$. The time
dependence of the state $|\psi\rangle$ can be read off from $
    |\psi,t\rangle
    =
    e^{-\imath H t}|\psi\rangle
$

The wavefunctional on this discretized space is given by
\[
    \Psi_k(\{\phi_{l,m,j}\}) = \frac{N_0}{\sqrt{2}}
    H_1(\sqrt{\omega_k} y_k)e^{-\frac{1}{2} y\Omega y}
\]
with $N_0 = \frac{\omega^{\frac{1}{4}}}{\sqrt{\pi^{\frac{1}{2}}}}$,
$y = S \phi$ where $S$ diagonalizes $K$, $(K)^{\alpha} = S^T
(\Omega)^{2 \alpha} S$ and $H_1(x) = 2x$ is a Hermite Polynomial. In
what follows it will be useful to compare our results to those
obtained by studying the ground state wavefunctional
\begin{equation}
\label{E:gswf}
    \Psi_{0}(\{\phi_{l,m,j}\}) = N_0 e^{-\frac{1}{2}y \Omega y}.
\end{equation}

We shall consider a general linear superposition of single excited
oscillators
\begin{align}
    \Psi &= \sum_k f_k \Psi_k\\
         & = \frac{N_0}{\sqrt{2}}
            \left(2 \phi S^T \Omega^{\frac{1}{2}} f\right)
            e^{-\frac{1}{2} \phi K^{\frac{1}{2}} \phi}.
\end{align}
For convenience, we define
\begin{equation}
\label{E:defofv}
    v_k = \left(f \Omega^{\frac{1}{2}} S\right)_k.
\end{equation}
Note that $v_n = A^{-\frac{1}{2}} \sum_j j f_j (e^{-\imath H
t})_{j,n}$. This allows us to write $\rho_{out}=\int \psi
\psi^{\star} d\phi_{in}$ as
\begin{equation}
\label{E:defofrhoout}
    \rho_{out}(w,w^{\prime}) =
        2\int (v \cdot \phi)(v^{\star} \cdot \phi^{\prime}) |N_0|^2
        e^{-\frac{1}{2} \phi K^{\frac{1}{2}} \phi}
        e^{-\frac{1}{2} \phi^{\prime} K^{\frac{1}{2}} \phi^{\prime}}
        d^n z,
\end{equation}
where we have used $\phi = \begin{pmatrix} z , w \end{pmatrix}$, and
$\phi^{\prime} = \begin{pmatrix} z , w^{\prime} \end{pmatrix}$.

The right-hand-side of eq.~(\ref{E:defofrhoout}) is a Gaussian
integral over the first $n$ coordinates. Using the notation
\begin{equation}
    v =
        \begin{pmatrix} v_1 \\ v_2 \end{pmatrix},
    \quad
\label{E:defofK12}
    K^{\frac{1}{2}} =
        \begin{pmatrix}
            A & B \\
            B^T & C
        \end{pmatrix},
\end{equation}
we find
\begin{align}
\notag
    \rho_{out}(w,w^{\prime}) =& 2 \rho_{0,out}(w,w^{\prime})
        \times \\
        \notag
        &\times
        \Bigg(
            \left(\frac{1}{2} v_1 A^{-1} B(w+w^{\prime})\right)
            \left(\frac{1}{2} v_1^{\star} A^{-1} B(w+w^{\prime})\right)
            +\frac{1}{2} v_1 A^{-1} v_1^{\star}-
            \\
            \notag
            &\phantom{\times\Bigg(}
            -(v_2 \cdot w) \frac{1}{2} v_1^{\star}A^{-1}
                B(w+w^{\prime})
            -(v_2^{\star} \cdot w^{\prime}) \frac{1}{2} v_1 A^{-1}
            B(w+w^{\prime})+
            \\
            &\phantom{\times\Bigg(}
            + (v_2 \cdot w)(v_2^{\star} \cdot w^{\prime})
        \Bigg),
\label{E:rhoin}
\end{align}
where
\begin{equation}
    \rho_{0,out}(w,w^{\prime})=|N_0|^2\frac{\pi^{\frac{n}{2}}}{|A|^{\frac{1}{2}}}
        \exp\left(\frac{1}{4} (w+w^{\prime})B^T A^{-1} B (w+w^\prime) -
        \frac{1}{2}(w C w+ w^{\prime} C w^{\prime})\right)
\end{equation}
is the density matrix corresponding to the vacuum state
(\ref{E:gswf}). One can check that $\text{Tr}\rho_{0,out}=
\text{Tr}\rho_{out} = 1$.

The eigenvalues of $\rho_{0,out}$ may be obtained numerically
\cite{Srednicki} (see also \cite{Plenio} for an analytic approach).
An approximation for the eigenvalues of the first excited state has
been discussed in \cite{Das}. Since we are interested in exhibiting
the nonunitary time evolution of the state measured by an external
observer at $r>r_s$, we shall evaluate $\text{Tr}(\rho_{out}^2)$; if
$\text{Tr}(\rho_{out}^2)$ is time dependent then, since
$\text{Tr}(\rho_{out})=1$, at least two of its eigenvalues are time
dependent.

To calculate $\text{Tr}(\rho_{out}^2)$ we carry out another Gaussian
integral. For the ground state we find
\begin{equation}
    \label{E:Trrhoin2ground}
    \text{Tr}(\rho_{0,out}^2)
        =\frac{|K|}{|A||\Upsilon|^{\frac{1}{2}}},
\end{equation}
where we have defined
\begin{equation}
\label{E:defofUp}
    \Upsilon =\begin{pmatrix}
                C-\frac{1}{2} B^T A^{-1}B & -\frac{1}{4}B^T A^{-1} B\\
                -\frac{1}{4}B^T A^{-1} B & C-\frac{1}{2} B^T A^{-1}B
            \end{pmatrix}.
\end{equation}
If $K$ were block diagonal such that $B=0$, then we would have had
$|\Upsilon| = |C|^2$, so that $\text{Tr}(\rho_{0,out}^2)=1$. We
expect this since $B=0$ implies that the `in' and `out'
oscillators are not entangled.

For our collapsing shell (\ref{E:psiini}) we find, after some
algebra, that
\begin{equation}
\label{E:Trrhoin2}
    \text{Tr}(\rho_{out}^2)=
        \left[\left( v \left[ \begin{pmatrix} A^{-1} & 0 \\ 0 & 0 \end{pmatrix} +
        U_B \right] v^{\star} \right)^2
        \\
        +\left( v U_A v^{\star} \right)^2
        +
        \left| v U_B v \right|^2
        \right]
    \text{Tr}(\rho_{0,out}^2)
\end{equation}
where
\begin{equation}
    U_{A/B} = \begin{pmatrix}
        \frac{1}{4} A^{-1} B(o_A+o_B)B^{T} A^{-1} &
            -\frac{1}{2} A^{-1} B(o_A+o_B) \\
        -\frac{1}{2} (o_A+o_B) B^T A^{-1} &
            o_{A/B}
    \end{pmatrix}
\end{equation}
and $o_A$ and $o_B$ are defined through
\begin{equation}
    \Upsilon^{-1} = \begin{pmatrix}
        o_A & o_B \\
        o_B & o_A
    \end{pmatrix}
\end{equation}
(where $\Upsilon$ is given in (\ref{E:defofUp}), and the block
matrices $A,B$ and $C$ are given in (\ref{E:defofK12})). As in the
ground state case, one may check that $\text{Tr}(\rho_{out}^2)=1$
if $B=0$.

In principle, the time dependence of $\rho_{out}$ can be extracted
from (\ref{E:Trrhoin2}). It is clear that if $f$ (recall $v=f
\Omega^{1/2} S$) has a single entry (implying that the state of the
system is an energy eigenstate) then $\text{Tr}(\rho_{out}^2)$ will
be time independent. The same applies to the ground state
$\text{Tr}(\rho_{0,out}^2)$ (\ref{E:Trrhoin2ground}). In fact, as
discussed earlier, any stationary state will generate a time
independent $\rho_{out}$, meaning that the time evolution will be
unitary. (Curiously, the same applies to coherent states (see
\cite{Das}).) To obtain the time dependence of
$\text{Tr}(\rho_{out}^2),$ we shall resort to numerical methods.

In figure~\ref{F:mainplot}, we have plotted the radial wavefunction
$\sum_j j f_j (e^{-\imath H t})_{j,n}$ as a function of the lattice
location, $n$, and time, $t$, for the time interval $[4a,15a]$.  We
used a radial lattice of size $N+1=50$. The initial wavefunction was
chosen such that at $t=0$ it is a Gaussian shell localized at $r_s =
33a$ with width $\sigma = \sqrt{5} a$, and momentum $-1a$. The thick
(red) line specifies the location of the maximum of the
wavefunction. Once the center of the wavefunction passes $r_s = 25
a$ (at $t=8.84 a$) we trace over the region $r<r_s$. In the figure,
this region is bounded by a semi-transparent membrane. The value of
$\text{Tr}(\rho_{out}^2)$ at $t>8.84 a$ is also plotted.

As the wavefunction enters the inaccessible region $r<25a$,
$\text{Tr}(\rho_{out}^2)$ starts increasing. This increase is to be
expected: once the wavefunction is `almost' completely inside the
region $r<r_s,$ then the one particle excitation of the density
matrix $\rho_{in}$ would describe an `almost' pure state implying
$\text{Tr}(\rho_{in}^2)/\text{Tr}(\rho_{0,in}^2)=1$. Since
$\rho_{in}$ and $\rho_{out}$ have equal eigenvalues (up to zeros),
we get the behavior depicted in the plot.

If the tracing procedure is carried out before $t=8.84 a$, then one
finds that $\text{Tr}(\rho_{out}^2)/\text{Tr}(\rho_{0,out}^2)$ drops
smoothly from $\sim 1$ (when the wavefunction is localized at
$r>r_s$) to $\sim 0.4$ at $t \sim 8.8 a$. Similarly, one can extend
the analysis to $t>15 a,$ whence the wavefunction ``bounces back"
from the origin: as the wavefunction leaves the $r>r_s$ region,
another sharp change in
$\text{Tr}(\rho_{out}^2)/\text{Tr}(\rho_{0,out}^2)$ is noted. At
very long time scales, the wavefunction will bounce back and forth
from the origin and IR boundary of space. In this case one will find
a semi-periodic behavior for $\text{Tr}(\rho_{out}^2)$. In general,
we observe that $\text{Tr}(\rho_{out}^2) = \alpha(t)
\text{Tr}(\rho_{0,out}^2)$ (see eq.~(\ref{E:Trrhoin2})), where
the coefficient $\alpha$ approaches unity when the wavefunction is
localized either in the $r<r_s$ region or when it is localized in
the $r>r_s$ region.

%\begin{figure}[btp]
%\begin{center}
\FIGURE{
\scalebox{0.81}{\includegraphics{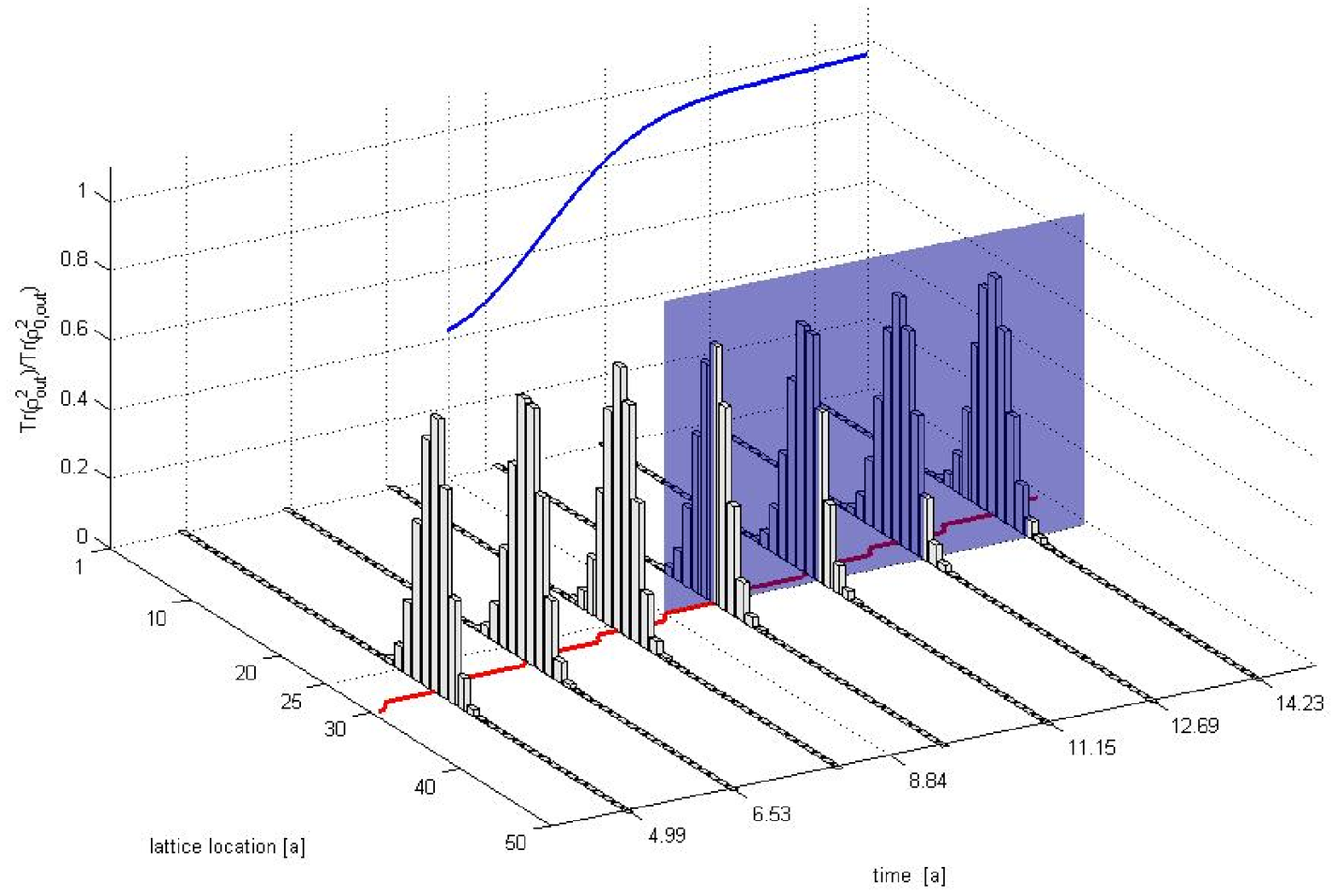}}%
\caption{\label{F:mainplot} A plot of the radial component of a
collapsing spherical wavefunction as a function of the lattice
location, $n$, and time, $t$. The radial lattice has $50$ points of
spacing $a$. At $t=0$ the wavefunction is a Gaussian shell localized
at $33a$ with width $\sigma = \sqrt{5} a$, and momentum $-1/a$. The
thick (red) line specifies the location of the maximum of the
wavefunction. Once the center of the wavefunction passes $r_s = 25
a$ (at $t=8.84 a$) we trace over the region $r<r_s$ (bounded by a
semi-transparent membrane).
$\frac{\text{Tr}(\rho_{out}^2)}{\text{Tr}(\rho_{0,out}^2)}$ is also
plotted for $t>8.84 a$.
}%
}
%\end{center}
%\end{figure}

We point out that since the wavefunctions corresponding to
$\phi_{0,0,j}|0\rangle$ are orthogonal under the Klein Gordon norm
then they are not orthogonal under the Dirac norm. Therefore, the
plot only gives a suggestive description of the region where the
wavefunction is localized. Also, since the scalar field is massless,
a zero momentum s-wave will split into an outgoing shell and an
ingoing shell. The amplitude of the outgoing shell is small
initially because the wavefunction has been given a large initial
ingoing momentum.

\section{Discussion}

The main point of this exercise is to emphasize that nonunitary time
evolution of a density matrix is an ubiquitous phenomenon and does
not signify a breakdown either of unitarity or of quantum mechanics.
Whenever an observer has a limited domain of observation, she will
describe observables in terms of a density matrix resulting from
tracing over states outside that domain \cite{feynman}.  If her
states are entangled with states beyond her domain, then she will
experience a density matrix corresponding to a mixed state.  It may
be static, or it may be time-dependent.  If time-dependent, it may
or may not be describable in terms of the dynamical variables
associated with her limited domain of observation, depending on the
situation.  Although the global state may be simple and involve
unitary evolution, the time dependence of the density matrix of a
subdomain can be quite complicated and appear nonunitary.

What implications might this have for gravitational collapse of
matter to a black hole?   Once a horizon forms, all observables
can be described in terms of a density matrix associated with the
states within a causal volume since, obviously, states beyond the
horizon are unobservable.   This situation is inherently
time-dependent due to black hole radiation. By energy
conservation, the horizon should shrink as this radiation is
emitted. In our view, the time dependence of the density matrix
describing the region outside the horizon is very likely to appear
to be nonunitary.  Globally, we expect that the evolution is
unitary and, if the matter started in a pure state, the total
entropy will remain zero. In this case, the apparent entropy of
the black hole would be entirely the result of entanglement.

One major difference between our setup and a collapsing black hole
is that our example is not singular and hence does not contain an
analog of what happens to matter as it falls into a region of high
curvature, where a classical singularity appears. Yet, if matter
does not disappear from the universe that remains when the black
hole has completely evaporated, one may speculate about the time
evolution of the black hole based on this example. On one hand, one
will find that from the point of view of a causal observer the
evolution of the black hole will seem nonunitary. On the other hand,
from a global viewpoint the wavefunction will evolve in a perfectly
unitary manner. The dichotomy between the two observers is only a
result of one of them restricting her observations to part of the
Hilbert space, much like the case of a system coupled to a heat
bath.  At the end of the day, when the black hole has evaporated,
there should be a unitary transformation from the state before the
horizon formed to the state after the horizon has disappeared.

%%% ------------------------------------------------------------------------
%\section*{Acknowledgments}
\acknowledgments%
We would like to thank Doron Cohen and Shanta de~Alwis for useful
discussions and Ehud Yarom for help with the figure.  The research
of M.~B.~E. was supported in part by the National Science
Foundation under Grant No. PHY99-07949. The research of A.~Y. was
supported in part by the German Science Foundation (DFG).

\bibliography{nonubib}

\providecommand{\href}[2]{#2}\begingroup\raggedright\begin{thebibliography}{10}

\bibitem{Srednicki}
M.~Srednicki, {\it Entropy and area},  {\em Phys. Rev. Lett.} {\bf 71} (1993)
  666--669, [\href{http://xxx.lanl.gov/abs/hep-th/9303048}{{\tt
  hep-th/9303048}}].

\bibitem{Bombelli}
L.~Bombelli, R.~K. Koul, J.-H. Lee, and R.~D. Sorkin, {\it A quantum source of
  entropy for black holes},  {\em Phys. Rev.} {\bf D34} (1986) 373.

\bibitem{Sol}
S.~N. Solodukhin, {\it Entanglement entropy and the ricci flow},
  \href{http://xxx.lanl.gov/abs/hep-th/0609045}{{\tt hep-th/0609045}}.

\bibitem{Casini1}
H.~Casini and M.~Huerta, {\it Universal terms for the entanglement entropy in
  2+1 dimensions},  \href{http://xxx.lanl.gov/abs/hep-th/0606256}{{\tt
  hep-th/0606256}}.

\bibitem{Fursaev1}
D.~V. Fursaev, {\it Entanglement entropy in critical phenomena and analogue
  models of quantum gravity},  {\em Phys. Rev.} {\bf D73} (2006) 124025,
  [\href{http://xxx.lanl.gov/abs/hep-th/0602134}{{\tt hep-th/0602134}}].

\bibitem{Das}
S.~Das and S.~Shankaranarayanan, {\it How robust is the entanglement entropy -
  area relation?},  {\em Phys. Rev.} {\bf D73} (2006) 121701,
  [\href{http://xxx.lanl.gov/abs/gr-qc/0511066}{{\tt gr-qc/0511066}}].

\bibitem{Myung}
Y.~S. Myung, {\it Entanglement system, casimir energy and black hole},  {\em
  Phys. Lett.} {\bf B636} (2006) 324--329,
  [\href{http://xxx.lanl.gov/abs/gr-qc/0511104}{{\tt gr-qc/0511104}}].

\bibitem{Buniy1}
R.~V. Buniy and S.~D.~H. Hsu, {\it Entanglement entropy, black holes and
  holography},  \href{http://xxx.lanl.gov/abs/hep-th/0510021}{{\tt
  hep-th/0510021}}.

\bibitem{Parikh}
M.~K. Parikh, {\it The volume of black holes},  {\em Phys. Rev.} {\bf D73}
  (2006) 124021, [\href{http://xxx.lanl.gov/abs/hep-th/0508108}{{\tt
  hep-th/0508108}}].

\bibitem{Dou}
D.~Dou and B.~Ydri, {\it Entanglement entropy on fuzzy spaces},
  \href{http://xxx.lanl.gov/abs/gr-qc/0605003}{{\tt gr-qc/0605003}}.

\bibitem{Plenio}
M.~B. Plenio, J.~Eisert, J.~Dreissig, and M.~Cramer, {\it Entropy,
  entanglement, and area: analytical results for harmonic lattice systems},
  {\em Phys. Rev. Lett.} {\bf 94} (2005) 060503,
  [\href{http://xxx.lanl.gov/abs/quant-ph/0405142}{{\tt quant-ph/0405142}}].

\bibitem{Areascaling}
A.~Yarom and R.~Brustein, {\it Area-scaling of quantum fluctuations},  {\em
  Nucl. Phys.} {\bf B709} (2005) 391--408,
  [\href{http://xxx.lanl.gov/abs/hep-th/0401081}{{\tt hep-th/0401081}}].

\bibitem{Casini2}
H.~Casini, {\it Geometric entropy, area, and strong subadditivity},  {\em
  Class. Quant. Grav.} {\bf 21} (2004) 2351--2378,
  [\href{http://xxx.lanl.gov/abs/hep-th/0312238}{{\tt hep-th/0312238}}].

\bibitem{Einhorn}
M.~B. Einhorn and M.~Mahato, {\it Beyond the horizon},  {\em Phys. Rev.} {\bf
  D73} (2006) 104035, [\href{http://xxx.lanl.gov/abs/gr-qc/0506020}{{\tt
  gr-qc/0506020}}].

\bibitem{Implications}
R.~Brustein, D.~H. Oaknin, and A.~Yarom, {\it Implications of area scaling of
  quantum fluctuations},  {\em Phys. Rev.} {\bf D70} (2004) 044043,
  [\href{http://xxx.lanl.gov/abs/hep-th/0310091}{{\tt hep-th/0310091}}].

\bibitem{coldbosons}
R.~Brustein and A.~Yarom, {\it Entanglement induced fluctuations of cold
  bosons},  \href{http://xxx.lanl.gov/abs/hep-th/0501058}{{\tt
  hep-th/0501058}}.

\bibitem{Holred}
R.~Brustein and A.~Yarom, {\it Holographic dimensional reduction from
  entanglement in minkowski space},  {\em JHEP} {\bf 01} (2005) 046,
  [\href{http://xxx.lanl.gov/abs/hep-th/0302186}{{\tt hep-th/0302186}}].

\bibitem{Mukohyama}
S.~Mukohyama, {\it Comments on entanglement entropy},  {\em Phys. Rev.} {\bf
  D58} (1998) 104023, [\href{http://xxx.lanl.gov/abs/gr-qc/9805039}{{\tt
  gr-qc/9805039}}].

\bibitem{Holzhey}
C.~Holzhey, F.~Larsen, and F.~Wilczek, {\it Geometric and renormalized entropy
  in conformal field theory},  {\em Nucl. Phys.} {\bf B424} (1994) 443--467,
  [\href{http://xxx.lanl.gov/abs/hep-th/9403108}{{\tt hep-th/9403108}}].

\bibitem{AYThesis}
A.~Yarom, {\em Entanglement and Horizons}.
\newblock PhD thesis, Ben Gurion University of the Negev, 2006.

\bibitem{Sorkin}
R.~D. Sorkin, {\it Ten theses on black hole entropy},  {\em Stud. Hist. Philos.
  Mod. Phys.} {\bf 36} (2005) 291--301,
  [\href{http://xxx.lanl.gov/abs/hep-th/0504037}{{\tt hep-th/0504037}}].

\bibitem{TandA}
R.~Brustein and A.~Yarom, {\it Thermodynamics and area in minkowski space: Heat
  capacity of entanglement},  {\em Phys. Rev.} {\bf D69} (2004) 064013,
  [\href{http://xxx.lanl.gov/abs/hep-th/0311029}{{\tt hep-th/0311029}}].

\bibitem{KabStr}
D.~Kabat and M.~J. Strassler, {\it A comment on entropy and area},  {\em Phys.
  Lett.} {\bf B329} (1994) 46--52,
  [\href{http://xxx.lanl.gov/abs/hep-th/9401125}{{\tt hep-th/9401125}}].

\bibitem{Israel}
W.~Israel, {\it Thermo field dynamics of black holes},  {\em Phys. Lett.} {\bf
  A57} (1976) 107--110.

\bibitem{Iorio}
A.~Iorio, G.~Lambiase, and G.~Vitiello, {\it Black hole entropy, entanglement,
  and holography},  \href{http://xxx.lanl.gov/abs/hep-th/0204034}{{\tt
  hep-th/0204034}}.

\bibitem{EntinST}
R.~Brustein, M.~B. Einhorn, and A.~Yarom, {\it Entanglement interpretation of
  black hole entropy in string theory},  {\em JHEP} {\bf 01} (2006) 098,
  [\href{http://xxx.lanl.gov/abs/hep-th/0508217}{{\tt hep-th/0508217}}].

\bibitem{MalStr}
J.~M. Maldacena and A.~Strominger, {\it Ads(3) black holes and a stringy
  exclusion principle},  {\em JHEP} {\bf 12} (1998) 005,
  [\href{http://xxx.lanl.gov/abs/hep-th/9804085}{{\tt hep-th/9804085}}].

\bibitem{HawMalStr}
S.~Hawking, J.~M. Maldacena, and A.~Strominger, {\it Desitter entropy, quantum
  entanglement and ads/cft},  {\em JHEP} {\bf 05} (2001) 001,
  [\href{http://xxx.lanl.gov/abs/hep-th/0002145}{{\tt hep-th/0002145}}].

\bibitem{LouMar}
J.~Louko and D.~Marolf, {\it Single-exterior black holes and the ads-cft
  conjecture},  {\em Phys. Rev.} {\bf D59} (1999) 066002,
  [\href{http://xxx.lanl.gov/abs/hep-th/9808081}{{\tt hep-th/9808081}}].

\bibitem{MarYar}
D.~Marolf and A.~Yarom, {\it Lodged in the throat: Internal infinities and
  ads/cft},  {\em JHEP} {\bf 01} (2006) 141,
  [\href{http://xxx.lanl.gov/abs/hep-th/0511225}{{\tt hep-th/0511225}}].

\bibitem{entAdSCFT1}
S.~Ryu and T.~Takayanagi, {\it Holographic derivation of entanglement entropy
  from ads/cft},  {\em Phys. Rev. Lett.} {\bf 96} (2006) 181602,
  [\href{http://xxx.lanl.gov/abs/hep-th/0603001}{{\tt hep-th/0603001}}].

\bibitem{entAdSCFT2}
S.~Ryu and T.~Takayanagi, {\it Aspects of holographic entanglement entropy},
  \href{http://xxx.lanl.gov/abs/hep-th/0605073}{{\tt hep-th/0605073}}.

\bibitem{entAdSCFT3}
Y.~Iwashita, T.~Kobayashi, T.~Shiromizu, and H.~Yoshino, {\it Holographic
  entanglement entropy of de sitter braneworld},
  \href{http://xxx.lanl.gov/abs/hep-th/0606027}{{\tt hep-th/0606027}}.

\bibitem{entAdSCFT4}
D.~V. Fursaev, {\it Proof of the holographic formula for entanglement entropy},
   \href{http://xxx.lanl.gov/abs/hep-th/0606184}{{\tt hep-th/0606184}}.

\bibitem{entAdSCFT5}
S.~N. Solodukhin, {\it Entanglement entropy of black holes and ads/cft
  correspondence},  \href{http://xxx.lanl.gov/abs/hep-th/0606205}{{\tt
  hep-th/0606205}}.

\bibitem{entAdSCFT6}
R.~Emparan, {\it Black hole entropy as entanglement entropy: A holographic
  derivation},  {\em JHEP} {\bf 06} (2006) 012,
  [\href{http://xxx.lanl.gov/abs/hep-th/0603081}{{\tt hep-th/0603081}}].

\bibitem{entAdSCFT7}
T.~Hirata and T.~Takayanagi, {\it Ads/cft and strong subadditivity of
  entanglement entropy},  \href{http://xxx.lanl.gov/abs/hep-th/0608213}{{\tt
  hep-th/0608213}}.

\bibitem{Bekenstein}
J.~D. Bekenstein, {\it Black holes and entropy},  {\em Phys. Rev.} {\bf D7}
  (1973) 2333--2346.

\bibitem{Hawkingnonu}
S.~W. Hawking, {\it Breakdown of predictability in gravitational collapse},
  {\em Phys. Rev.} {\bf D14} (1976) 2460--2473.

\bibitem{Mav1}
J.~Bernabeu, J.~Ellis, N.~E. Mavromatos, D.~V. Nanopoulos, and
  J.~Papavassiliou, {\it Cpt and quantum mechanics tests with kaons},
  \href{http://xxx.lanl.gov/abs/hep-ph/0607322}{{\tt hep-ph/0607322}}.

\bibitem{Mav2}
J.~Bernabeu, N.~E. Mavromatos, and S.~Sarkar, {\it Decoherence induced cpt
  violation and entangled neutral mesons},
  \href{http://xxx.lanl.gov/abs/hep-th/0606137}{{\tt hep-th/0606137}}.

\bibitem{Mav3}
N.~E. Mavromatos and S.~Sarkar, {\it Methods of approaching decoherence in the
  flavour sector due to space-time foam},
  \href{http://xxx.lanl.gov/abs/hep-ph/0606048}{{\tt hep-ph/0606048}}.

\bibitem{Giddings1}
S.~B. Giddings, {\it Black hole information, unitarity, and nonlocality},
  \href{http://xxx.lanl.gov/abs/hep-th/0605196}{{\tt hep-th/0605196}}.

\bibitem{Giddings2}
S.~B. Giddings, {\it Locality in quantum gravity and string theory},
  \href{http://xxx.lanl.gov/abs/hep-th/0604072}{{\tt hep-th/0604072}}.

\bibitem{Vijay}
V.~Balasubramanian, D.~Marolf, and M.~Rozali, {\it Information recovery from
  black holes},  \href{http://xxx.lanl.gov/abs/hep-th/0604045}{{\tt
  hep-th/0604045}}.

\bibitem{Horowitz2}
G.~T. Horowitz and E.~Silverstein, {\it The inside story: Quasilocal tachyons
  and black holes},  {\em Phys. Rev.} {\bf D73} (2006) 064016,
  [\href{http://xxx.lanl.gov/abs/hep-th/0601032}{{\tt hep-th/0601032}}].

\bibitem{Einhorntalk}
M.~B. Einhorn, {\it The black hole information paradox},
  \href{http://xxx.lanl.gov/abs/hep-th/0510148}{{\tt hep-th/0510148}}.

\bibitem{Horowitz1}
G.~T. Horowitz and J.~M. Maldacena, {\it The black hole final state},  {\em
  JHEP} {\bf 02} (2004) 008,
  [\href{http://xxx.lanl.gov/abs/hep-th/0310281}{{\tt hep-th/0310281}}].

\bibitem{Hawkingrecent}
S.~W. Hawking, {\it Information loss in black holes},  {\em Phys. Rev.} {\bf
  D72} (2005) 084013, [\href{http://xxx.lanl.gov/abs/hep-th/0507171}{{\tt
  hep-th/0507171}}].

\bibitem{Hsu}
S.~D.~H. Hsu, {\it Spacetime topology change and black hole information},
  \href{http://xxx.lanl.gov/abs/hep-th/0608175}{{\tt hep-th/0608175}}.

\bibitem{dissipation}
U.~Weiss, {\em Quantum Dissipative systems}.
\newblock World Scientific Publishing Co. Pte. Ltd., P.O.Box 128, Farrer Road,
  Singapore 912805, 1999.

\bibitem{Nakajima}
S.~Nakajima, {\it On quantum theory of transport phenomena--diffusion},  {\em
  Prog.\ Theor.\ Phys.} {\bf 20} (1958) 948.

\bibitem{Zwanzig}
R.~Zwanzig, {\it Ensemble method in the theory of irreversibility},  {\em J.\
  Chem.\ Phys.} {\bf 33} (1960) 1338.

\bibitem{boulder}
R.~Zwanzig, {\em Lectures in Theoretical Physics}, vol.~3.
\newblock W. E. Brittin, W. B. Downs, J. Downs (eds.), Interscience, 1961.

\bibitem{Cardy}
P.~Calabrese and J.~L. Cardy, {\it Evolution of entanglement entropy in
  one-dimensional systems},  {\em J. Stat. Mech.} {\bf 0504} (2005) P010,
  [\href{http://xxx.lanl.gov/abs/cond-mat/0503393}{{\tt cond-mat/0503393}}].

\bibitem{feynman}
R.~P. Feynman, {\em Statistical mechanics, 2nd ed.}
\newblock Perseus Books, Advanced Book Program, Reading, MA, 1998.

\end{thebibliography}\endgroup

\end{document}